\documentclass[5p]{elsarticle} 

\usepackage{hyperref}
\usepackage{atlasphysics}
\usepackage{subfigure}
\usepackage{xspace}
\usepackage{lineno}
\usepackage{multirow}

\newcommand{\mT}{\ensuremath{m_{\mathrm{T}}}}
\def\Htotautau{\ensuremath{A/H/h \to \tau^+\tau^-}\xspace}
\def\Ztoee{\ensuremath{Z/\gamma^* \to e^+e^-}\xspace}
\def\Ztomumu{\ensuremath{Z/\gamma^* \to \mu^+\mu^-}\xspace}
\def\Ztotautau{\ensuremath{Z/\gamma^* \to \tau^+\tau^-}\xspace}
\def\MET{\ensuremath{E_{\mathrm{T}}^{\mathrm{miss}}}\xspace}
\def\METx{\ensuremath{E_{\mathrm{x}}^{\mathrm{miss}}}\,}
\def\METy{\ensuremath{E_{\mathrm{y}}^{\mathrm{miss}}}\,}
\def\mVis{\ensuremath{m_{\tau\tau}^{\mathrm{visible}}}\xspace}
\def\mT{\ensuremath{m_{\mathrm{T}}}\xspace}
\def\pt{\ensuremath{p_{\mathrm{T}}}\xspace}
\def\Wjets{\ensuremath{W+\mathrm{jets}}\xspace}
\def\Zjets{\ensuremath{Z/\gamma^*+\mathrm{jets}}\xspace}
\def\ttbar{\ensuremath{t\bar{t}}\xspace}
\def\rqcd{\ensuremath{r_{\mathrm OS/SS}^{\mathrm{QCD}}}\xspace}

\def\rwjets{\ensuremath{r_{\mathrm{OS/SS}}^{W}}\xspace}

\def\nOS{\ensuremath{n_{\mathrm{OS}}^{Bkg}}\xspace}

\def\nSSall{\ensuremath{n_{\mathrm{SS}}^{Bkg}}\xspace}

\def\nSSwjets{\ensuremath{n_{\mathrm{SS}}^{W}}\xspace}

\def\nOSSSqcd{\ensuremath{n_{\mathrm{OS-SS}}^{\mathrm{QCD}}}\xspace}
\def\nOSSSwjets{\ensuremath{n_{\mathrm{OS-SS}}^{W}}\xspace}
\def\nOSSSzjets{\ensuremath{n_{\mathrm{OS-SS}}^{Z}}\xspace}
\def\nOSSSother{\ensuremath{n_{\mathrm{OS-SS}}^{\mathrm{other}}}\xspace}
\def\meff{\ensuremath{m_{\tau\tau}^{\mathrm{effective}}}\,}
\def\currentlumi{$36~\ipb$}
\def\currentlumiprecise{$36.1\pm1.2~\ipb$}

\begin{document}
\begin{frontmatter}

\title{\vspace{-0.5cm} \flushright{\normalsize{CERN-PH-EP-2011-104}} \\ \center{
Search for neutral MSSM Higgs bosons decaying to $\tau^+\tau^-$ pairs 
in proton-proton collisions at $\sqrt{s}=7$~TeV with the ATLAS detector}}
  \author{The ATLAS Collaboration}

\begin{abstract}
A search for neutral Higgs bosons decaying to pairs of $\tau$ leptons 
with the ATLAS detector at the LHC is presented. The analysis is based 
on proton-proton collisions at a center-of-mass energy of $7\tev$,  
recorded in 2010 and corresponding to an integrated luminosity of \currentlumi.
After signal selection, $276$ events are observed in this data sample.
The observed number of events is consistent with the total expected 
background of $269 \pm 36$ events. Exclusion limits at the $95\%$ 
confidence level are derived for the production cross section of a generic 
Higgs boson $\phi$ as a function of the Higgs boson mass and 
for $A/H/h$ production in the Minimal Supersymmetric Standard Model (MSSM)
as a function of the parameters $m_A$ and $\tan\beta$.
\end{abstract}
     
	
\end{frontmatter}

\section{Introduction}
\label{sec:introduction}

Discovering the mechanism responsible for electroweak symmetry breaking
and the origin of mass for elementary particles is one of the major
goals of the physics program at the Large Hadron Collider~(LHC)~\cite{LHC}.
In the Standard Model (SM) this mechanism requires the existence of a 
scalar particle, the Higgs boson~\cite{ENGLERT,HIGGS,HIGGS2,HIGGS3,Guralnik:1964eu}. 
In extensions of the Standard Model to the Minimal Supersymmetric Standard Model 
(MSSM)~\cite{MSSM1,MSSM2}, two Higgs doublets of opposite hypercharge are required, 
resulting in five observable Higgs bosons. Three of these 
are electrically neutral ($h$, $H$, and $A$)
while two are charged ($H^\pm$). At tree level 
their properties such as masses, widths, and branching ratios can be
predicted in terms of only two parameters, often chosen to be the
mass of the $CP$-odd Higgs boson, $m_A$, and the ratio of
the vacuum expectation values of the two Higgs doublets, $\tan\beta$.
The Higgs boson production proceeds mainly via gluon fusion
or in association with $b$~quarks, with the latter becoming
more important for large $\tan\beta$. 

In this paper, a search for neutral MSSM Higgs bosons in the decay mode 
\Htotautau with the ATLAS detector~\cite{ATLASDetector} is presented. 
The decay into a $\tau^+\tau^-$ pair is a promising channel 
since the coupling of the Higgs bosons to third-generation fermions is 
strongly enhanced over large regions of the MSSM parameter space.
The search considers Higgs boson decays to 
$e\mu4\nu$,
$e\tau_{had}3\nu$, and
$\mu\tau_{had}3\nu$, 
where  $\tau_{had}$ denotes a hadronically decaying $\tau$ lepton. 
These topologies have branching ratios of $6\%$,  $23\%$, and $23\%$, respectively. 
This analysis is complementary to previous searches at 
the $e^+e^-$ collider LEP at CERN~\cite{LEPLimits} 
and similar to those performed at the $p\bar{p}$ collider
Tevatron at Fermilab~\cite{TevatronLimits,Abazov:2011qz}, and
extends to regions of the MSSM parameter space untested by these machines.
The CMS Collaboration has recently published results of a
similar analysis~\cite{CMSLimit}.

\section{Event samples} 
\label{sec:data_mc}

The data used in this search were recorded with the ATLAS detector 
in proton-proton collisions at a center-of-mass energy of $\sqrt{s}=7$~TeV
during the 2010 LHC run. 
The ATLAS detector is described in detail elsewhere~\cite{ATLASDetector}.
In the ATLAS coordinate system, 
polar angles~$\theta$ are measured with respect to the LHC beamline and
azimuthal angles~$\phi$ are measured in the plane transverse to the beamline.
Pseudorapidities $\eta$ are defined as $\eta = - \ln \tan \frac{\theta}{2}$.
Transverse momenta are computed from the three-momenta $p$ as
$p_T = |p| \ \sin\theta$.
The integrated luminosity of the data sample,
considering only data-taking periods where all relevant detector subsystems 
were fully operational, is \currentlumiprecise~\cite{lumi}.
The data were collected using a single-electron trigger with \pt\ threshold
in the range $10-15\gev$ for the $e\tau_{had}$ and $e\mu$ final states,
and a single-muon trigger with \pt\ threshold in the range $10-13\gev$
for the $\mu\tau_{had}$ final state. With respect to the signal selection 
described below, the total trigger efficiencies are 
$99\%$ and $82\%$  for electrons and muons, respectively. 
Events that pass the trigger are selected if they have a reconstructed 
vertex that is formed by three or more tracks and lies within $15$~cm of 
the nominal interaction point along the beam axis. 

The cross sections for Higgs boson production have been calculated using
{\sc HIGLU}~\cite{Spira:1995mt} 
and ggh@nnlo~\cite{ggh_at_nnlo} for the gluon fusion process. 
For the $b$-quark associated production, 
a matching scheme described in~\cite{SantanderMatching} is used to combine  
the next-to-leading order (NLO) calculation for $gg \rightarrow b\bar{b}A/H/h$ in the 4-flavor 
scheme~\cite{Dittmaier:2003ej,Dawson:2005vi} and 
the next-to-next-to-leading order (NNLO) calculation for $b\bar{b} \rightarrow A/H/h$ in the 5-flavor 
scheme~\cite{Harlander:2003ai}. 
The masses, couplings, and branching ratios
of the Higgs bosons are computed with FeynHiggs~\cite{FeynHiggs}. 
The ratio of the MSSM Yukawa couplings and their SM values have been used to 
derive the MSSM cross sections from the respective SM cross sections.
Details of the calculations and associated $\alpha_{S}$, parton distribution 
function (PDF) and scale uncertainties can be found in
Ref.~\cite{LHCHiggsCrossSectionWorkingGroup:2011ti}.
The direct $gg\rightarrow A/H/h$~production is simulated with 
MC@NLO~\cite{MCatNLO}, and the associated $b\bar{b}A/H/h$ production with 
SHERPA~\cite{SHERPA}.
Both $gg \to A$ and $bbA$ samples are generated 
at 11 values of $m_A$, in the range from $90$ to $300\gev$.
These samples are also used for the $H$ and $h$ bosons assuming 
the same kinematics of the decay products. 
For any given $m_A$ and $\tan\beta$, the masses $m_H$ and $m_h$ of the $H$ and
$h$ bosons are calculated in the $m_h^{max}$ MSSM benchmark scenario~\cite{MSSMmhmax} 
and $A$ boson events with $m_A$ closest to $m_H$ and $m_h$, respectively,
are added to these samples with appropriately scaled cross sections to obtain
a signal sample for $A/H/h$ production.
The increase of the Higgs boson natural width
with $\tan\beta$ is neglected as it is small compared with the 
experimental mass resolution.
Table~\ref{tab:MCxsec} shows the signal cross section times branching ratio
for $\tan\beta=40$ and $m_A = 120 \gev$.

\begin{table}[htdp]
\begin{center}
\caption{Cross sections for signal and background processes. 
For $A/H/h$ production, the cross section is multiplied by the branching 
ratio for $A/H/h \to \tau^+\tau^-$.
The signal cross sections are given for $\tan\beta=40$ and the three 
values quoted correspond to $A/H/h$ production, respectively.
For $m_A=120\gev$ and $\tan\beta=40$, the $H$ and $h$ boson masses
in the $m_h^{max}$ scenario are $m_H=131.0\gev$ and $m_h=119.5\gev$.
}
\label{tab:MCxsec}
\begin{tabular}{lc}
\\
\hline \hline
Signal process                                           & $\sigma$ $\times$ BR [pb] \\ \hline

$bbA/H/h     (\to\tau\tau)$, $m_A = 120\gev$  & $30.8/1.08/32.3$ \\
$gg\to A/H/h (\to\tau\tau)$, $m_A= 120\gev$   & $20.5/2.97/19.3$ \\
\hline
Background process                                           & $\sigma$ [pb] \\ 
\hline
$W\rightarrow \ell$+jets ($\ell=e, \mu, \tau$) & $10.46 \times 10^3$ \\
$Z/\gamma^*\rightarrow \ell^+\ell^-$+jets ($m_{\ell\ell}>10\gev$)& $4.96 \times 10^3$ \\
$t\bar{t}$                                  & $164.6$ \\
Single-top ($t-, s-$ and $Wt$-channels)         & $58.7$, $3.9$, $13.1$ \\
Di-boson ($WW$, $WZ$ and $ZZ$)                 & $46.2$, $18.0$, $5.6$ \\ 
\hline \hline

\end{tabular}
\end{center}
\end{table}

Processes producing $W$ or $Z$ bosons that subsequently decay into leptons 
constitute the most important background. 
These processes include
\Wjets, \Zjets, where $\gamma^*$ denotes a virtual photon, top-quark (\ttbar and single-top) and electroweak
di-boson ($WW$, $WZ$, $ZZ$) production.
Here, 
$Z/\gamma^*\to\tau^+\tau^- + \rm jets$ events 
constitute a largely irreducible background for Higgs boson masses close to 
the $Z$ boson mass.
 $Z/\gamma^* \to \ell^+\ell^- + \rm jets$ ($\ell = e,\mu$) events
 contribute if one of the charged leptons or an accompanying jet is 
 misidentified. 
 Due to its large cross section, jet production in Quantum Chromodynamics (QCD) processes 
 provides a significant background contribution if there are real leptons 
 from decays of heavy quarks or if jets are misidentified as 
 electrons, muons, or hadronic $\tau$ decays.

The production of $W$ and $Z$ bosons in association with jets is simulated 
with the ALPGEN~\cite{Alpgen} and PYTHIA~\cite{Pythia} generators. 
The \ttbar\ and single-top processes are generated with MC@NLO, and 
for di-boson production HERWIG~\cite{Herwig} and MC@NLO are used. 
The loop-induced process $gg \to WW$ is generated with gg2WW~\cite{GG2WW}.
For events generated with ALPGEN, HERWIG,
MC@NLO and gg2WW the parton shower and hadronization are simulated with 
HERWIG and the underlying event with JIMMY~\cite{JIMMY}.
The programs TAUOLA~\cite{TAUOLA,TAUOLA2} and PHOTOS~\cite{PHOTOS} are used to 
model the decays of $\tau$ leptons and the radiation of photons,
respectively, in all event samples except those generated with SHERPA.

Table~\ref{tab:MCxsec} summarizes the inclusive cross sections for the 
above processes, which are used to normalize the simulated event samples.
The cross section for single gauge boson production is calculated at NNLO
in QCD perturbation theory~\cite{Gavin:2010az}, for $t\bar{t}$ production at
NLO and next-to-leading logarithms (NLL)~\cite{Moch:2008ai,Langenfeld:2009tc}, 
and for single-top and di-boson production at NLO \cite{MCatNLO}.
No simulated samples for the QCD jet background are used, as this background
is entirely estimated with data. 
All simulated samples are processed through a full simulation 
of the ATLAS detector based on GEANT4~\cite{Geant4,ATLASSIM}. 
To match the pile-up (overlap of several interactions 
in the same bunch crossing) observed in the data, minimum-bias events 
~\cite{pileup1,pileup2} are overlaid to the generated signal and 
background events, and the resulting events are reweighted so that
the distribution of the number of reconstructed vertices
per bunch crossing agrees with the data.

\section{Object reconstruction}
\label{sec:objects}

Electron candidates are reconstructed from a cluster of energy deposits in the 
electromagnetic calorimeter matched to a track in the inner detector. 
The cluster must have a shower profile consistent with 
an electromagnetic shower~\cite{Aad:2010yt}.
Electron candidates are required to have a transverse momentum above $20\gev$ 
and a pseudorapidity in the range $|\eta|<1.37$ or $1.52<|\eta|<2.47$.
Muon candidates are reconstructed by combining tracks in the muon spectrometer 
with tracks in the inner detector~\cite{Aad:2010yt}.
They must have a transverse momentum above $10$~GeV and 
a pseudorapidity in the range $|\eta| < 2.5$ and $<2.4$ in the
$\ell\tau_{had}$ and $e\mu$ final states, respectively.
Isolation requirements are imposed on electron (muon) candidates by requiring that the 
additional transverse energy in the calorimeter cells in a cone
of radius $\Delta R = \sqrt{(\Delta\eta)^2+(\Delta\phi)^2} = 0.3 \,(0.4)$ 
centered on the lepton direction is less than $10\%$ ($6\%$) of the electron 
(muon) transverse energy (momentum). 
In addition, the sum of the transverse momenta of all tracks with $p_T > 1\gev$
within $\Delta R=0.4$ around the lepton direction, excluding the lepton track,  
must be less than 6\% of the lepton track transverse momentum.
The reconstruction of candidates for hadronic $\tau$ decays
is based on calorimeter jets reconstructed with
the anti-$k_T$ algorithm~\cite{AntiKT,AntiKT1} with a distance parameter $R = 0.4$, 
seeded using three-dimensional topological calorimeter energy clusters.
Their identification, including vetoing electrons and muons, 
is based on observables that describe the shape of the 
calorimeter shower and on tracking information, which are combined in 
a likelihood discriminator~\cite{ATLASTauSimpleID}. 
A $\tau$ candidate must have a visible transverse momentum, $p_T^{\tau,vis}$, above $20$~GeV, 
a pseudorapidity in the range $|\eta|<2.5$, $1$ or $3$ associated tracks 
($p_T > 1\gev$) and a total charge of $\pm 1$, computed from all tracks associated with 
the candidate.
The efficiency of the $\tau$ identification 
for 1-prong (3-prong) $\tau$ candidates with $p_T^{\tau,vis}>20 \gev$ 
is about $65\%$ ($60\%$) and the probability to misidentify a jet 
as a $\tau$ lepton, as determined from a di-jet control sample, is 
about $10\%$ ($5\%$).
When candidates fulfilling the above criteria overlap with each other geometrically~(within $\Delta R<0.2$), 
only one of them is selected. The overlap is resolved by selecting muons, electrons and $\tau$ candidates 
in this order of priority. The missing transverse momentum in the event, $\MET=\sqrt{(\METx)^2+(\METy)^2}$, 
is reconstructed as the vector sum
of all topological calorimeter energy clusters in the region $|\eta|<4.5$
and corrected for identified muons~\cite{Aad:2010yt}.

 \section{Event selection}
\label{sec:selection}

The signatures of $\Htotautau \to e \mu 4\nu$ signal events are one 
isolated electron, one isolated muon and \MET due to the undetected neutrinos
from the two $\tau$ decays. 
Exactly one electron with $p_\mathrm{T}^{e} > 20 \gev$ and one muon with 
$p_\mathrm{T}^{\mu} > 10 \gev$ with opposite electric charge
are required. 
In order to suppress backgrounds from \ttbar, 
single-top and di-boson production two additional requirements are applied.
The scalar sum of the transverse momentum of the electron, the transverse
momentum of the muon and the missing transverse momentum must be smaller than $120 \gev$, and the 
azimuthal opening angle between the electron and the muon must be larger 
than $2.0$~rad.

The signatures of \Htotautau signal events, where one $\tau$ lepton decays leptonically and
the other hadronically, are an isolated electron or muon, $\ell$, 
a $\tau$ candidate, $\tau_{had}$, and \MET due to the undetected neutrinos from the two $\tau$ 
decays. 
Exactly one electron or muon with $p_\mathrm{T}^{e} > 20 \gev$ or $p_\mathrm{T}^{\mu} > 15 \gev$ 
and one oppositely-charged $\tau$ candidate with  $p_\mathrm{T}^{\tau, vis} > 20 \gev$ are 
required in the event. Events with more than one electron or muon, using the lepton $\pt$ 
thresholds 
from the object definition given in Section~\ref{sec:objects}, 
are rejected to suppress events from $Z/\gamma^* \to\ell^+\ell^-$ $(\ell = e,\mu)$ 
decays and from \ttbar and single-top production.
A missing transverse momentum above $20\gev$ is required to reject events 
with jets from QCD processes as well as 
$Z/\gamma^*\to\ell^+\ell^-$ $(\ell = e,\mu)$ decays.  
Events with real leptons from $W\to\ell\nu$ decays are suppressed
by requiring the transverse mass of the \mbox{$\ell$-\MET} system, defined as
\begin{eqnarray}
\mT =
\sqrt{2p_{\mathrm{T}}^{\ell}E_{\mathrm{T}}^{\mathrm{miss}}(1-\cos\Delta\phi)},
\end{eqnarray}
to be below $30\gev$.
Here $p_{\mathrm{T}}^{\ell}$ is the transverse momentum of the electron or muon and 
$\Delta\phi$ is the angle between the electron or muon and the \MET vector in the plane perpendicular to the beam direction.

Table~\ref{tab:cutflow} compares the number of selected events
in data with those expected from the simulation of various background processes,
not including QCD jet production.
After the full selection,  $70$, $74$, and $132$ data events are observed in
the $e\mu$, $e\tau_{had }$, and $\mu\tau_{had }$ channels, respectively.
The estimation of backgrounds based on data control samples 
used for the final results of the analysis is discussed in
Section~\ref{sec:BackgroundEstimation}.
The signal efficiency amounts to $7 (3) \%$ for $m_A=120\gev$ and $9 (8)\% $ 
for $m_A=200\gev$ in the $e\mu \ (\ell \tau_{had})$ final states.

\begin{table*}[bt]
\begin{center}
\caption{Number of selected events in data and expected from Monte Carlo (MC) simulation 
         for a data sample corresponding to \currentlumi.
         The total $A/H/h$ signal yields for $m_A=120\gev$ and $\tan\beta=40$ are shown in the rightmost column.
         Only the MC statistical uncertainties are quoted.
         No MC expectation is given for the QCD jet background because
         this background can only be reliably estimated with data 
         (it amounts to $2.1^{+3.1}_{-2.1}$ and $7.8 \pm 7.0$ events for the $e\mu$ and 
          the combined $e\tau_{had}$ and $\mu\tau_{had}$ final states, 
          respectively, as described in Sections~\ref{sec:QCDll} and~\ref{sec:QCDandWjets}).
}
\label{tab:cutflow}
\vspace*{0.3cm}
\begin{tabular}{c|c|c|ccccc|c} \hline \hline
                &Data &Total MC bkg  & $W$+jets      & Di-boson      & $t\bar{t} +$&$Z/\gamma^{*} \to$&$Z/\gamma^{*}\to$&Signal\ ($m_A=120\gev$, \\
                &     &(w/o QCD)     &               &               & single-top  &$ee,\mu\mu$       &$\tau^+\tau^-$   & $\tan\beta=40$)\\ \hline
$e\mu$          & 70  &60.4$\pm$1.2  & 0.7$\pm$0.5   & 2.8$\pm$0.1   & 2.5$\pm$0.1 & 0.8$\pm$ 0.1     & 53.5$\pm$1.0    & 16.0$\pm$0.3\\
$e\tau_{had}$   & 74  & 72.8$\pm$2.7 & 25.1$\pm$1.8  & 0.37$\pm$0.02 & 4.1$\pm$0.2 &10.6$\pm$1.0      & 32.6$\pm$1.8    & 18.7$\pm$0.5\\
$\mu\tau_{had}$ & 132 &145.2$\pm$3.9 &   41.5$\pm$2.1& 0.59$\pm$0.03 & 5.9$\pm$0.2 &  11.5$\pm$1.1    & 85.7$\pm$3.1  &   36.6$\pm$0.8\\
\hline \hline
\end{tabular}
\end{center}
\end{table*}

After the selection of signal candidates in the 
$e\mu$ final state, the effective mass, \meff, is used as the discriminating variable
to search for a potential Higgs boson signal.
Here, \meff\ is calculated as the invariant mass of the electron, muon
and \MET system according to
\begin{eqnarray}
\meff = \sqrt{(p_e+p_\mu+p_{\mathrm{miss}})^2},
\end{eqnarray}
where $p_e$ and $p_\mu$ denote the four-vectors of the electron and muon,
respectively,
and the missing momentum four-vector is defined by
$p_{\mathrm{miss}} = (\MET, \METx, \METy, 0)$.

In the $e\tau_{had}$ and $\mu\tau_{had}$ final states, the visible $\tau^+\tau^-$ mass, \mVis, defined 
as the invariant mass of 
the electron or muon from the leptonic $\tau$ decay and the hadron(s) from the hadronic $\tau$ 
decay, is used as the discriminating variable. 

\begin{figure}[bt]
  \centering
  \subfigure{
    \includegraphics[width=0.33\textwidth]{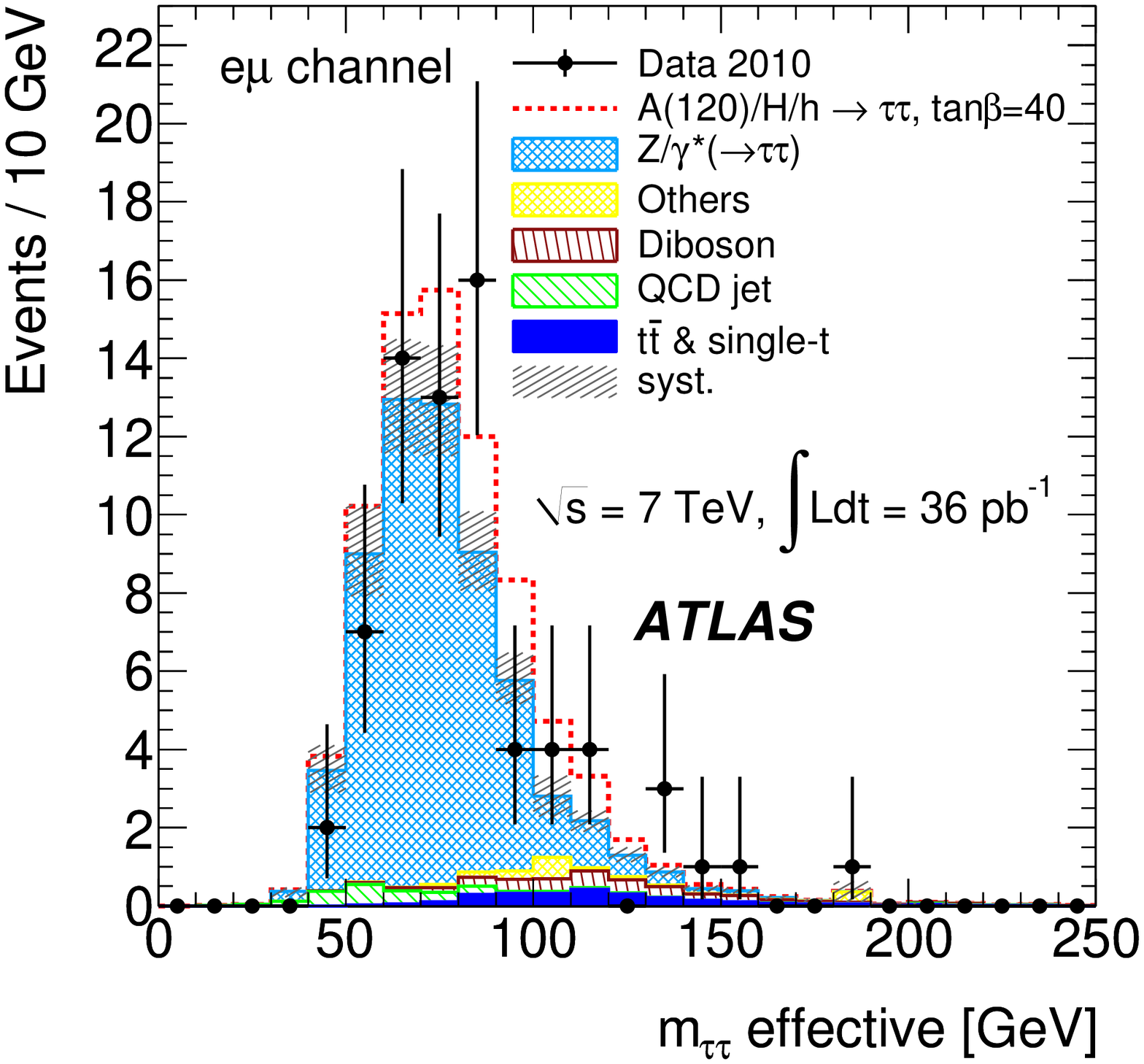}
  }
  \subfigure{
    \includegraphics[width=0.33\textwidth]{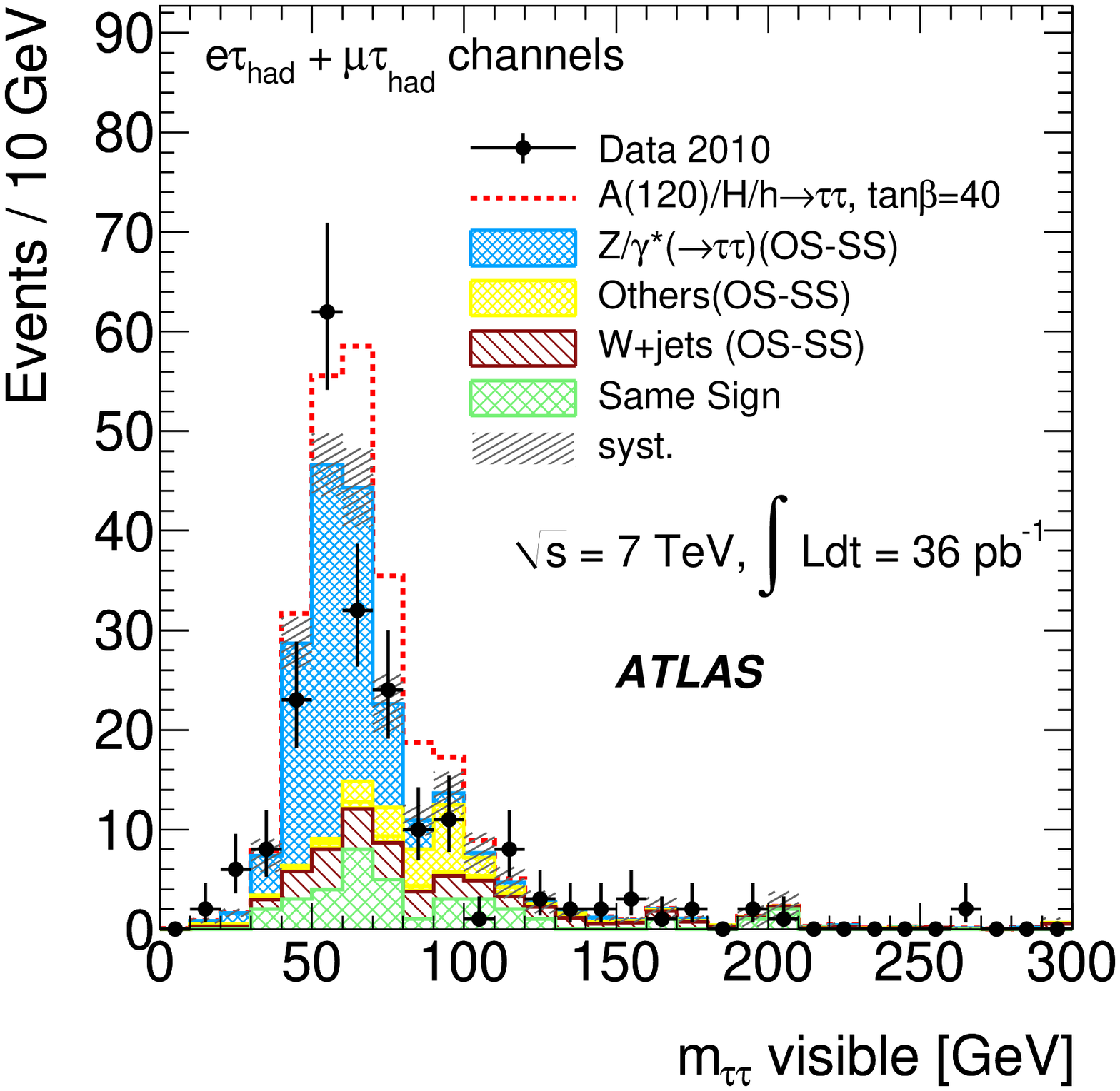}
  }
\vspace*{-0.3cm}

  \caption{Effective mass distribution for the $e\mu$ final state (top)
  and visible mass distribution for $\ell\tau_{had}$ final states (bottom).
  The data are compared with the background expectation
  and an added hypothetical signal. 
  ``OS-SS'' denotes the difference between 
  the opposite-sign and same-sign event yields. Further explanations are given 
  in the text.
}
  \label{fig:MvisMeff}
\end{figure}

Figure~\ref{fig:MvisMeff} shows distributions of \meff 
and \mVis for the data, compared to the background expectations described 
in Section~\ref{sec:BackgroundEstimation}.

\section{Background estimation} 
\label{sec:BackgroundEstimation}
In the search for a Higgs boson signal 
the normalization and shape of the \mVis and \meff\ 
distributions
for the sum of all background contributions
have to be determined.
Data control samples are used, where possible, to estimate or validate the most relevant 
background sources: 
\Ztotautau and QCD jet production in the $e\mu$ final state, and
\Wjets, \Ztotautau, and QCD jet production in the $\ell \tau_{had}$ final state. 
The remaining backgrounds given in Table~\ref{tab:cutflow} are estimated solely from simulation.

\subsection{QCD jet background in the $e \mu$ final state}
\label{sec:QCDll}
For the estimation of the QCD jet background, four independent samples
are selected by using selection criteria on two variables:
the isolation of the electron and muon and their charge product.
The signal region A is defined by the
selection criteria defined above, {\it i.e.} opposite-sign isolated leptons.
Region B contains same-sign isolated leptons, region C opposite-sign
anti-isolated leptons, and region D same-sign anti-isolated leptons.
Anti-isolated leptons are obtained by inverting the isolation criteria
described in Section~\ref{sec:objects}.
The shape of the \meff distribution in the signal region A is taken from 
control region~C and the normalization is derived by
$n_A= r_{C/D} \times n_B$. Here, $n_A$ and $n_B$ denote the event yields in regions 
A and B and $r_{C/D}$ the ratio of the event yields in regions C and D after
subtracting the contribution from non-QCD jet backgrounds estimated from simulation.
This method relies on the assumption that the two variables used to define the four 
regions are uncorrelated and that the shape of the \meff distribution does not 
depend on the isolation or charge product requirement. This has been verified
by comparing the event yields and shapes of the \meff\ distribution in
data for regions C and D and in further control regions defined by
the requirement of one isolated and one non-isolated lepton.

After subtracting the contribution from non-QCD jet backgrounds, estimated from simulation,
the QCD jet event yield in region B is found to be $n_B = 1.07\pm 1.57 (stat.)$ 
and the ratio $r_{C/D}$ is determined to be
$r_{C/D}= 1.97 \pm 0.12 (stat.)$. The QCD jet event yield in the
signal region is therefore estimated to be $n_A=2.1 ^{+3.1}_{-2.1} (stat.)$. 
Systematic uncertainties are discussed in Section~\ref{sec:systematics}.

\subsection{Background in the $\ell\tau_{had}$ final states}
\label{sec:QCDandWjets}
The method to estimate the QCD and \Wjets backgrounds~\cite{BkgCONF} 
is based on both data and simulation and uses events with same-sign charges
of the electron/muon and the $\tau_{had}$ candidate. 
It relies on the assumptions that the shape of the \mVis distribution 
for these backgrounds is the same for opposite-sign (OS) and same-sign (SS) events 
and that their ratio is the same in the signal region, defined by the nominal selection,
and in background-enhanced QCD and \Wjets control regions.
These assumptions have been verified with simulated events. 
The method is referred to as the baseline method
and is used to derive the results for the $\ell\tau_{had}$ channel.
It is cross-checked with an alternative background estimation method.

The total number of opposite-sign background events 
in the signal region, \nOS, can be expressed as
\begin{equation}
\label{eq:nOS_start}
  \nOS = \nSSall + \nOSSSqcd + \nOSSSwjets + \nOSSSzjets + \nOSSSother,
\end{equation}
where $\nSSall$ is the sum of all same-sign backgrounds in the signal region
and the remaining terms are the differences between opposite-sign and same-sign
events for the QCD, \Wjets, \Ztotautau, and other backgrounds. 
The ratio of opposite-sign and same-sign events for the QCD background, \rqcd, 
is expected to be  close to unity.
For \Wjets, a significant deviation of the ratio \rwjets from unity is expected 
since \Wjets production is dominated by $gu$/$gd$-processes that often give rise
to a jet originating from a quark whose charge is anti-correlated with the $W$ charge. 
From simulation, the ratio $\rwjets = 2.24 \pm 0.13 (stat.)$ is obtained.

Using $\nOSSSwjets = (\rwjets-1) \cdot \nSSwjets$ and assuming $\rqcd=1$,
Eq.~\ref{eq:nOS_start} can be approximated by
\begin{equation}
\label{eq:nOS_final}
  \nOS=\nSSall + (\rwjets-1) \cdot \nSSwjets + \nOSSSzjets + \nOSSSother.
\end{equation}
Each of the terms in Eq.~\ref{eq:nOS_final} is estimated separately and 
for each bin in the \mVis distribution, 
thus not only an estimation of the background normalization 
but also of the \mVis shape is obtained. 
The total number of same-sign events \nSSall\ is
determined for the nominal selection 
except for changing the opposite-sign charge requirement to same-sign. 
In the full \mVis range, $36$ same-sign events are selected in data.
The contributions from \Ztotautau and other backgrounds are taken from simulation:
$\nOSSSzjets =112 \pm 4 (stat.)$ 
and
$\nOSSSother=26 \pm 2 (stat.)$.
The \Wjets term in Eq.~\ref{eq:nOS_final} is estimated to be
$(\rwjets-1) \cdot \nSSwjets = 31 \pm 2 (stat)$.
Here, the number of same-sign \Wjets events in the signal region, \nSSwjets, and the ratio \rwjets 
are determined in a \Wjets-dominated data control region 
selected by replacing the $\mT<30\gev$ requirement in the nominal selection 
by  $\mT>50\gev$. The small contribution from backgrounds other than \Wjets 
is subtracted based on simulation. 
A value of $\rwjets = 2.41 \pm 0.15 (stat.)$ is obtained.
It has been checked in simulation that this ratio is 
approximately independent of 
the \mT range and can thus be used for the signal region. 
\nSSwjets is obtained by scaling the number of events in the \Wjets control region by
the ratio of events in the signal and control regions determined from simulation.
The shape of the \mVis distribution for this contribution 
is taken from simulation.

The assumption $\rqcd \approx 1$ used in Eq.~\ref{eq:nOS_final} is checked   
with a data control sample that is dominated by relatively low-$\ET$ jets 
from QCD processes, as expected in the signal region. 
This sample is selected by replacing the requirement $\MET>20$~GeV with $\MET<15$~GeV and
relaxing the isolation of the electron/muon candidate.
After subtraction of the other backgrounds using simulation,  
a value of $\rqcd = 1.16\pm 0.04 (stat.)$ is obtained.
The observed deviation of \rqcd from unity is taken into account in the 
determination of systematic uncertainties for the final result,
leading to a total systematic uncertainty of $19\%$ on \rqcd.
This uncertainty also includes an uncertainty associated with
the dependence of \rqcd on the lepton isolation and detector effects. 

The total background estimate obtained from Eq.~\ref{eq:nOS_final} 
is $\nOS = 206 \pm 7 (stat.)$, to be compared
with $206$ events observed in data.

An alternative background estimation is performed, which
provides separate estimates of the QCD and \Wjets background contributions
and is used to
cross-check the results of the baseline method discussed before.
For the QCD jet background the same method and assumptions
as described in Section~\ref{sec:QCDll} for the $e\mu$ final state are used, but
replacing one of the leptons ($e$ or $\mu$) by the $\tau_{had}$ candidate 
and using the \mVis distribution instead of \meff.
The shape of the \mVis distribution is taken from region~B and scaled by
the ratio of event yields in regions C and D: 
$r_{C/D} =  1.12 \pm 0.04 (stat.)$. The resulting estimate of the 
QCD jet background in the signal region is 
$n_A^{\rm QCD} = r_{C/D} \times n_B =  7.8 \pm 7.0 (stat.)$.
The estimate of the \Wjets background is obtained by deriving a scale 
factor of $0.83 \pm 0.04 (stat.)$ for the normalization of the 
simulated \mVis distribution in a 
$W$-dominated data control sample. This control region is defined by replacing 
the $\mT < 30\GeV$ requirement in the nominal selection by $70<\mT<120\GeV$.  
The shape of the \Wjets background is taken from simulation. The estimated 
number of \Wjets events for the nominal selection amounts 
to $54.8 \pm 2.1 (stat.)$ events.
Adding the expected number of events for \Ztotautau and the other backgrounds from simulation 
(see Table~\ref{tab:cutflow})
to the sum of the estimated QCD jet and \Wjets yields, a total background contribution of $211 \pm 8 (stat.)$ events
is obtained, which agrees well with the 206 events observed in data.  
The \mVis shapes predicted by the two methods are found to agree as well.

\begin{table*}[b!h!t!]
\begin{center}
\caption{Uncertainties on the number of selected events 
for those background contributions 
that are at least partially estimated from simulation and for a hypothetical signal ($m_A=120\gev$).
All numbers are given in \%. When two numbers are given the first refers to the $e\mu$ final state
and the second to the $\ell\tau_{had}$ final states. If an uncertainty does not apply for a certain background,
this is indicated by a ``-''.
\label{tab:systematics}}

\vspace*{0.3cm}

\begin{tabular}{c|ccccc|c} \hline \hline
             & $W$+jets & Di-boson& $t\bar{t} +$&$Z/\gamma^{*} \to$&$Z/\gamma^{*}\to$&Signal \ ($m_A=120\gev$, \\
             &          &         & single-top  &$ee,\mu\mu$ &$\tau^+\tau^-$   & $\tan\beta=$40)   \\ \hline
$\sigma_{theory}$    &  5/-   & 7       &  10        &  5       &  5         & 14  \\
Acceptance        &  3/-   & 1/2     &   5/2      &  2/14    &  3/14      & 5/4-7 \\
$ e $  efficiency &  8/-   & 7/3     &   7/3      &  9/2     &  8/7       & 7/6   \\
$\mu$  efficiency &  2/-   & 2/2     &   2/2      &  2/2     &  2/2       & 2/2   \\
$\tau$ efficiency and fake rate &  -/-   & -/4     &   -/4      &  -/20    &  -/4       & -/4   \\
Energy scales and resolutions &  2/-   & 2/2     &   4/2      &  2/28    &  2/36      & 1/19  \\
Luminosity              & 3.4/-  &  3.4    &   3.4      &   3.4    &   3.4      & 3.4  \\
Total uncertainty & 11/-   &  11/9   &  15/12     &  11/38   &  11/40     & 16/25  \\
\hline \hline
\end{tabular}
\end{center}
\end{table*}
 
\subsection{Validation of the \Ztotautau background shape}
\label{sec:embedding}

\begin{figure}
  \centering
  \subfigure{
    \includegraphics[width=0.33\textwidth]{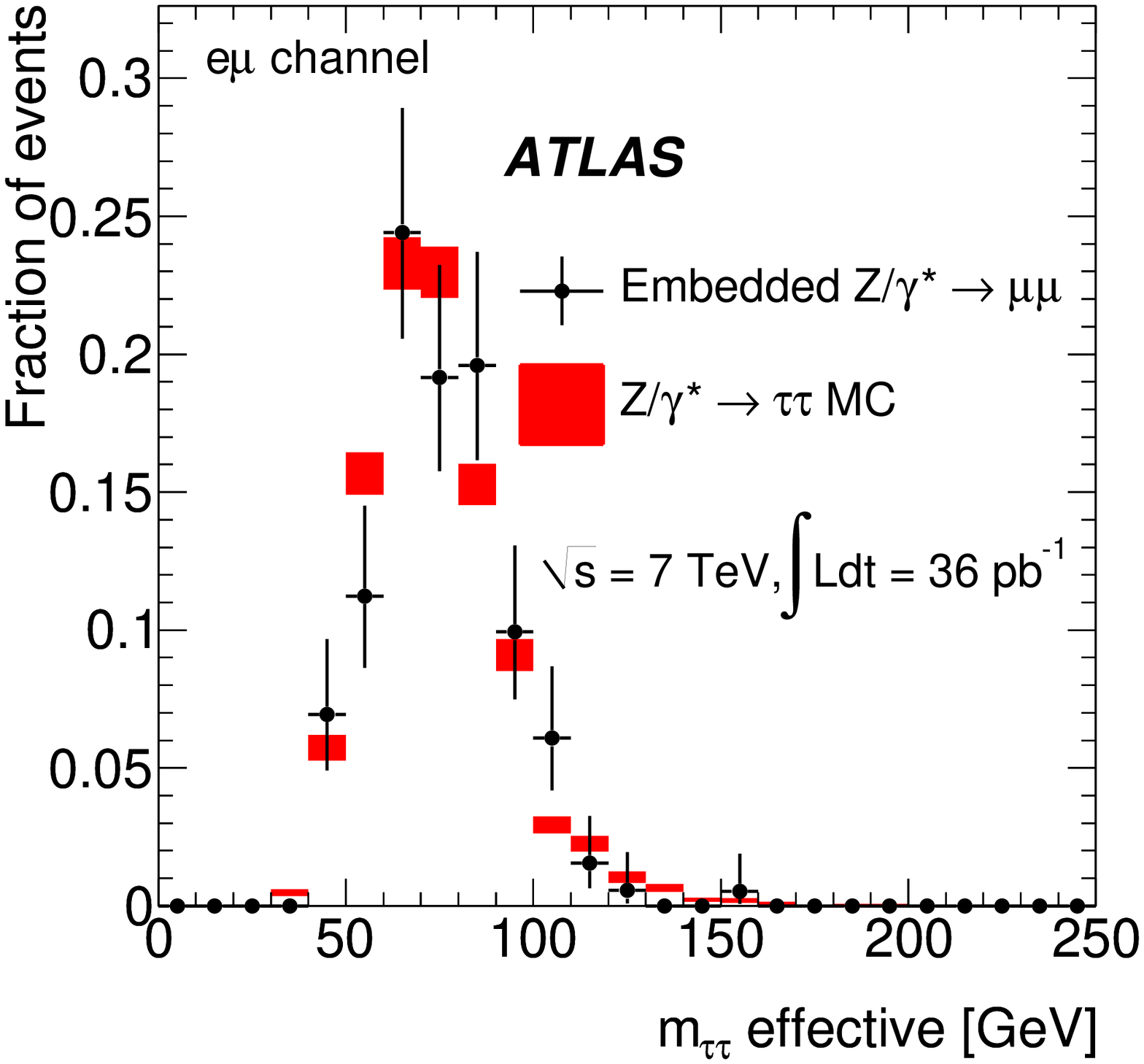}
  }
  \subfigure{
    \includegraphics[width=0.33\textwidth]{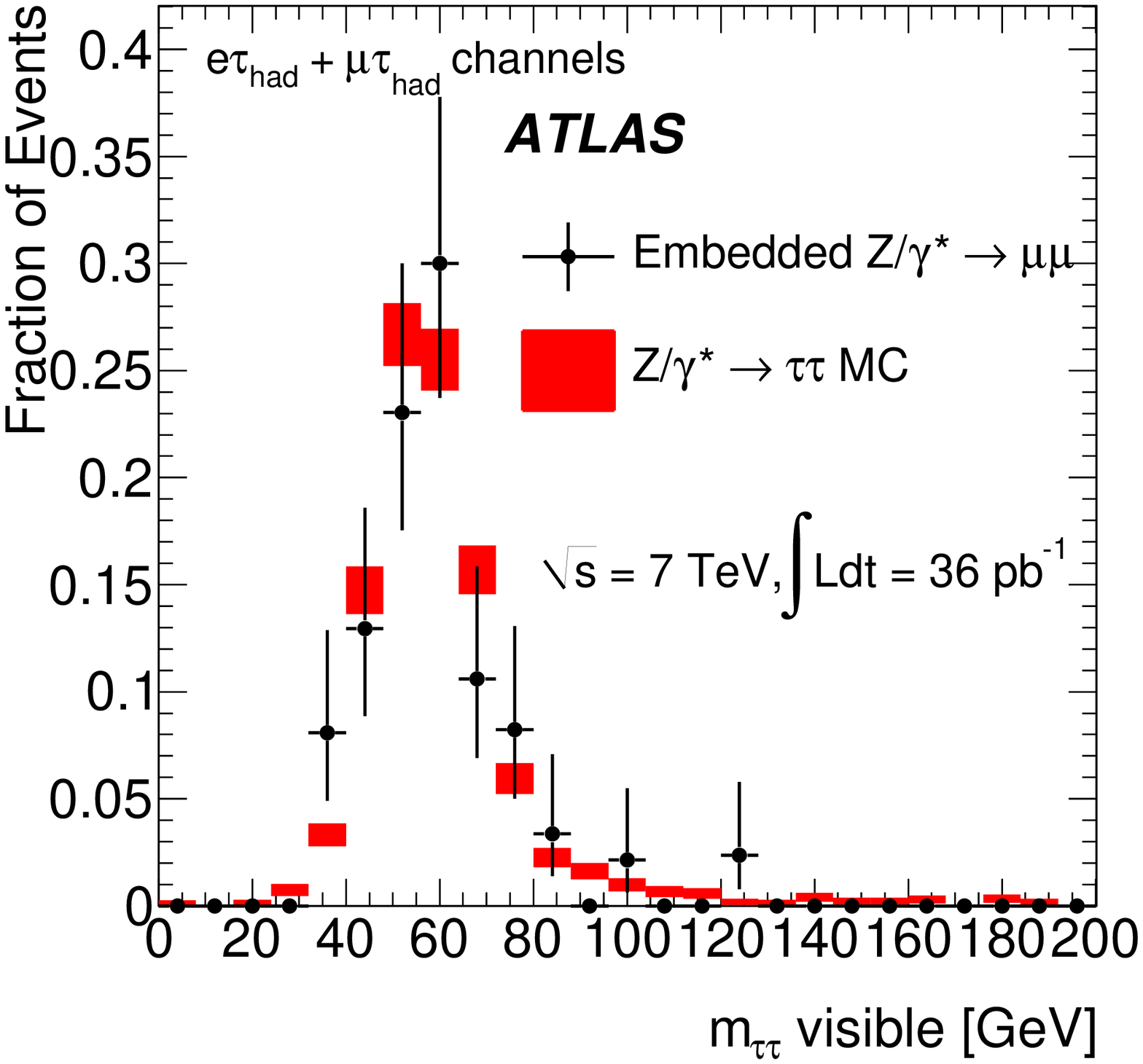}
  }
\vspace*{-0.3cm}

  \caption{Effective mass distribution for the $e\mu$ final state (top)
  and visible mass distribution for the $\ell\tau_{had}$ final states (bottom)
  for simulated \Ztotautau events (boxes) and $\tau$-embedded \Ztomumu events (points) 
  passing the signal selection. The size of the boxes and the length of the error bars 
  indicate the statistical uncertainty on the simulated and $\tau$-embedded sample, respectively.}
  \label{fig:embedding}
\end{figure}

The shape of the \mVis and \meff\ distributions for the irreducible \Ztotautau background
can be determined from a high-purity data sample of \Ztomumu events
in which the muons are removed and replaced by simulated $\tau$ leptons.
Thus, only the $\tau$ decays and the corresponding detector response
are taken from simulation, whereas the underlying $Z/\gamma^*$ kinematics and
all other properties of the event are obtained from the \Ztomumu data.
Figure~\ref{fig:embedding} compares the \mVis and \meff\ distributions of the
$\tau$-embedded sample with simulated \Ztotautau events. A good agreement
is observed within the sizable statistical uncertainties, 
justifying the use of the simulation for the determination of the \Ztotautau 
background. This background is normalized
according to the theoretical cross section in Table~\ref{tab:MCxsec}, which
agrees with the ATLAS $Z/\gamma^*\to\ell^+\ell^-$ cross section measurement~\cite{Aad:2010yt}.

 \section{Systematic uncertainties} 
\label{sec:systematics}
Systematic effects on the signal efficiency and the estimated number of
background events are evaluated. The uncertainties can be grouped in four
categories: theoretical inclusive cross sections, acceptance, knowledge of
detector performance and systematic uncertainties of the  data-driven approaches to estimate
the background contribution. 

The uncertainty on the theoretical inclusive cross section for each individual signal and background process 
is obtained from variations of the renormalization and factorization scales ($\mu_R$, $\mu_F$) 
by factors $1/2$ and $2$ and a variation 
of the strong coupling constant and the PDF sets within their uncertainties.
The uncertainty on the acceptance is estimated by varying $\mu_R$, $\mu_F$,
matching parameters in ALPGEN and the choice of the PDF in the generation of simulated
event samples. 
The uncertainty on the trigger efficiencies for electrons and muons is $1\%$.
The uncertainties due to the limited knowledge of the detector performance
are evaluated by varying the trigger, reconstruction and identification efficiencies
for electrons, muons and $\tau$ candidates, and by varying the energy resolution and energy scale
of electrons, muons, $\tau$ candidates, and energy deposits outside of these objects.
These are propagated in a fully correlated way into the \MET scale and resolution.
For the probability to misidentify electrons as $\tau$ candidates, a $20\%$ uncertainty
is assumed, resulting in a $20\%$ uncertainty on the \Ztoee background.

The size of the uncertainties from the different sources on the various background
processes which are at least partially estimated from simulated events are summarized
in Table~\ref{tab:systematics}. 
The luminosity uncertainty is $3.4\%$.

The dominant systematic uncertainty in the $\ell\tau_{had}$ final states is due to the 
variation of the jet and $\tau$ energy scales, which are dependent on transverse momentum
and pseudorapidity, by typically $7\%$ and $5\%$, respectively.
The difference in the impact of the energy scale and resolution
uncertainty on the expected event yields in the $\ell\tau_{had}$ and $e\mu$ final 
states is caused by requiring a hadronic $\tau$ decay with $p_T^{\tau,vis}>20\gev$ 
and a lower threshold $\MET >20\gev$ 
in the $\ell\tau_{had}$ final states, whereas in the $e\mu$ final state only 
an upper threshold of $p_\mathrm{T}^{e} + p_\mathrm{T}^{\mu} + \MET < 120 \gev$ is required.
The uncertainties, apart from the ones related to the data-driven techniques,
are treated as fully correlated between the three final states. 

The systematic uncertainty from the data-driven estimate
of the QCD jet background in the $e\mu$ final state corresponds to $0.8$ events.
It includes the systematic uncertainty on the subtracted non-QCD background 
($0.2$ events) and on the assumption of identical  \meff~ shapes in the different 
control regions (uncertainty on $r_{C/D}$ of  $0.78$). The final estimate for the 
QCD jet yield in the signal region is therefore
$n_A=2.1^{+3.1}_{-2.1} (stat.) \pm 0.8 (syst.) = 2.1^{+3.2}_{-2.1}$. 
The total uncertainty is dominated by the small event yield in control region B.

For the $\ell\tau_{had}$ channels, the most important uncertainties for the 
data-driven estimation of the QCD jet and \Wjets backgrounds (see Eq.~\ref{eq:nOS_final}) 
are the statistical uncertainty on the number of same-sign events in the signal region~(17\%) and 
the uncertainty on the ratios  $\rqcd$~(19\%) and \rwjets~(11\%).
An additional uncertainty of 10\% is derived from the \mT dependence 
of \rwjets, {\it i.e.} for the extrapolation from control to signal region.
The final estimate for the total background yield is
$\nOS = 206 \pm 7 (stat.) \pm 34 (syst.) = 206 \pm 35$.

The impact of the energy scale uncertainties of the electron, muon, $\tau$ candidate, and \MET 
on the shapes of the discriminating mass variables are included as an
additional correlated uncertainty in the derivation of the Higgs boson exclusion limits in 
Section~\ref{sec:limit}. All other systematic uncertainties have no significant effect on the mass shape. 

Combining the estimated contribution from the various background processes and their uncertainties 
results in the final background estimate shown in
Table~\ref{tab:final_background}.

\begin{table}[htdp]
\begin{center}
\caption{Observed numbers of events in data, for an integrated luminosity 
of \currentlumi, and total expected background contributions for the 
final states considered in this analysis, with 
their combined statistical and systematic uncertainties.}
\vspace*{0.3cm}
\label{tab:final_background}
\begin{tabular}{lcc} \hline \hline
Final state      & Exp. Background & Data \\ \hline
$e\mu$           &  63 $\pm$  7    &  70 \\  
$\ell\tau_{had}$  & 206 $\pm$ 35    & 206 \\  \hline
Sum              & 269 $\pm$ 36    & 276 \\
\hline \hline
\end{tabular}
\end{center}
\end{table}

\section{Results}
\label{sec:limit}
No significant excess of events is observed in the data, compared to the SM expectation.
Exclusion limits at the $95\%$ confidence level are 
set on the production cross section times branching ratio of a generic Higgs boson $\phi$ as a 
function of its mass and for MSSM Higgs boson $A/H/h$ production as a function of the parameters  
$m_A$ and $\tan\beta$. The exclusion limits are derived with the profile likelihood method~\cite{CCGV}
from an analysis of the \meff\ distribution for the 
$e\mu$  final state and the \mVis distribution for the 
$\ell\tau_{had}$ final states.

Systematic uncertainties are separated into common, fully correlated
(energy scale, acceptance, luminosity) and channel-specific, and 
are included as nuisance parameters. 
The  \meff and \mVis shape uncertainties due to variation of the energy scales 
of leptons and  \met\ for the backgrounds obtained from simulation are taken into account.

The $p$-values for the consistency of the observed data with the background-only hypothesis 
range from $3\%$ for a mass of 300 GeV to $59\%$
for a mass of 110 GeV for the combination of the $e\mu$ and $\ell\tau_{had}$ channels.

Background-only toy MC experiments are generated to find the median expected limit 
along with the $\pm 1\sigma$ and $+2\sigma$ error bands. 
As a protection against excluding the signal hypothesis 
in cases of downward fluctuations of the background, the 
observed limit is not allowed to fluctuate below $-1\sigma$ 
of the expected limit, {\it i.e.} a power-constrained limit (PCL,~\cite{PCL}), 
with the power required to be larger than 16\%, is given.

\begin{figure*}
  \centering

  \subfigure{
    \includegraphics[width=0.37\textwidth]{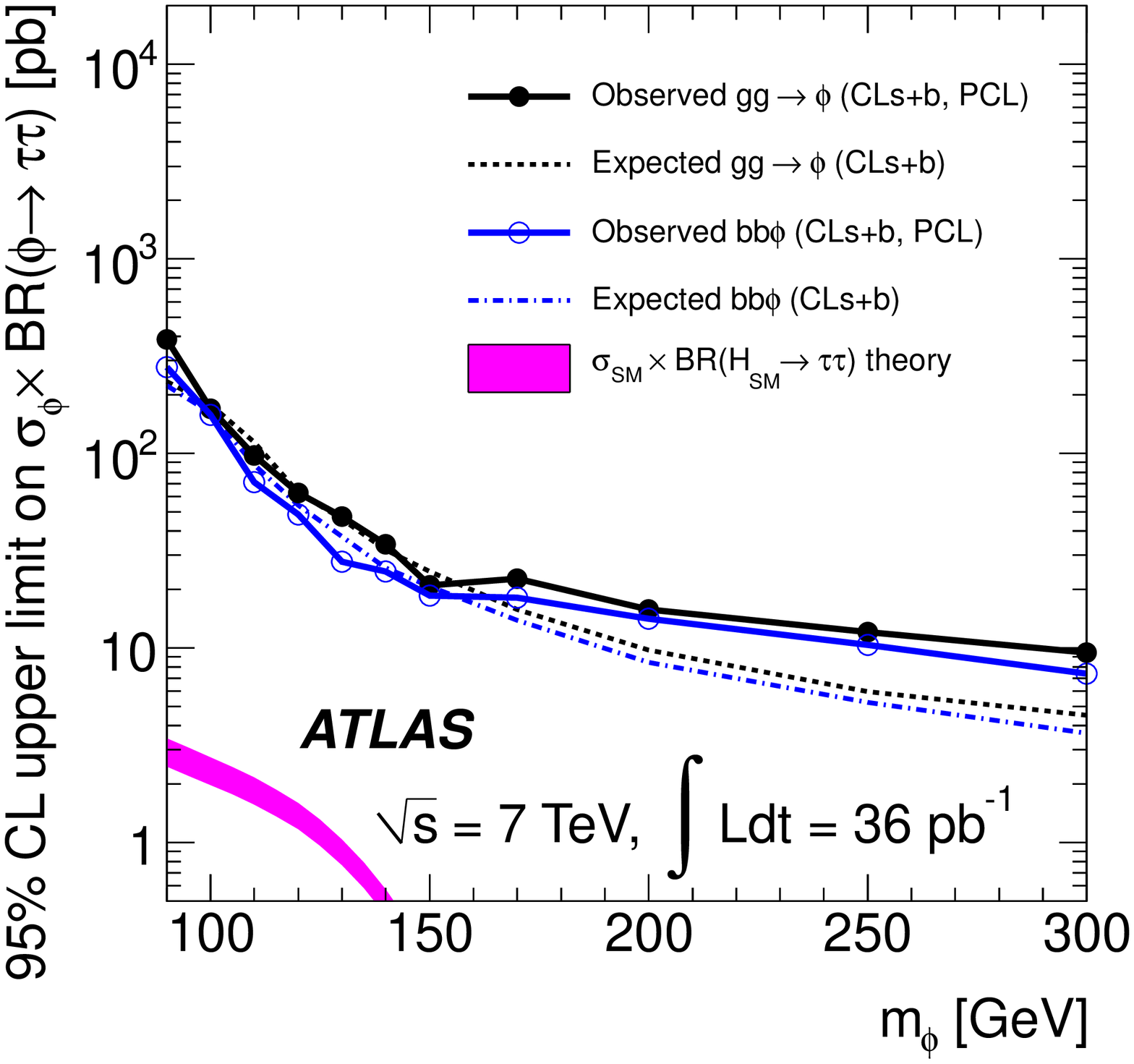}
  }
  \subfigure{
    \includegraphics[width=0.37\textwidth]{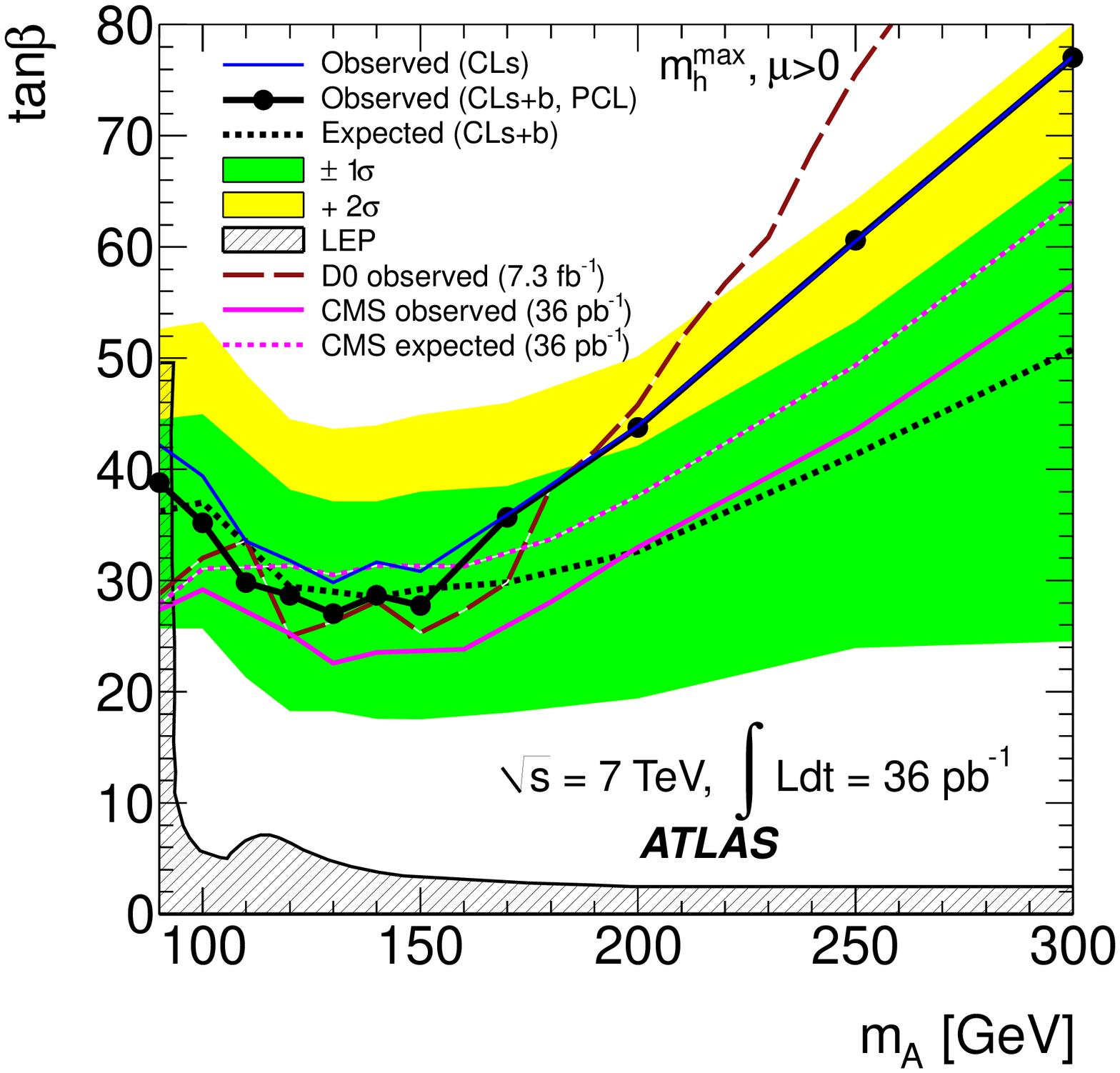}
  }

\vspace*{-0.3cm}

\caption{
Left: Expected and observed limits on the production cross section and 
branching ratio for a generic Higgs boson $\phi$, 
$\sigma_\phi \times BR(\phi\to\tautau)$,  at the $95\%$ confidence level, as a function of the 
      Higgs boson mass for both production modes considered.
      The solid and dashed lines show the observed and expected exclusion limits, respectively.
      For comparison the SM cross section, $\sigma_{SM} \times BR(H_{SM}\to\tautau)$, is also shown.
 Right: Expected and observed exclusion limits in the $m_A-\tan\beta$ plane of the MSSM.
        The region above the drawn limit curve is excluded at the $95\%$ confidence level.
        The dark grey (green) and light grey (yellow) bands correspond to the $\pm 1\sigma$ and
        $+ 2\sigma$ error bands, respectively.
        For comparison the observed limit based on $CL_{\rm s}$~\cite{Junk:1999kv,cls1} is shown 
        in addition to the one based on $CL_{\rm s+b}$.
        The observed limit is shown up to $\tan\beta \approx 80$ although it should be noted that the region
        $\tan\beta>65$ is considered to be theoretically not well under control~\cite{MSSM_Djouadi}.
        The exclusion limits from LEP, D0, and CMS are also shown.} 
  \label{fig:limit}
\end{figure*}

Figure~\ref{fig:limit} shows the resulting exclusion limits. 
The cross section limit is evaluated for signal acceptances of two 
different production processes, $gg \to \phi$ and b-quark associated 
production, where $\phi$ denotes a generic neutral Higgs boson.
Differences in the observed limits for the two
processes are small compared to the $1\sigma$ error and occur due to 
differences in the signal shapes used in the extraction of the limits.
The limit on the production cross section times branching ratio
into a pair of $\tau$ leptons for a generic Higgs boson $\phi$ 
is in the range between approximately  $300$~pb for a Higgs boson mass 
of 90 GeV and approximately $10$~pb for a Higgs boson mass of 300 GeV, with a small
dependence on the production mode considered.
The limit on the production of neutral MSSM Higgs bosons $A/H/h$ in the $\tan\beta-m_A$ plane,
also shown in Figure~\ref{fig:limit},
uses the $m_h^{max}$ scenario and Higgsino mass parameter $\mu>0$. 

\section{Conclusions}

In this paper, a search for neutral MSSM Higgs bosons $A/H/h$ with the ATLAS detector
in proton-proton collisions corresponding to an integrated luminosity of \currentlumi\ 
at a center-of-mass energy of $7 \tev$ is presented. 
Candidates for $A/H/h \to \tau^+\tau^-$ decays are selected
in the three final states $e\mu$, $e\tau_{had}$, and $\mu\tau_{had}$.
No evidence for a Higgs boson signal is observed in the reconstructed mass spectra.
Exclusion limits on both, the cross section for the production of a generic Higgs boson $\phi$
as a function of its mass and on MSSM Higgs boson production $A/H/h$
as a function of $m_A$ and $\tan\beta$, are derived. These results exclude regions
of parameters space beyond the existing limits from previous experiments at
LEP and the Tevatron and are similar to those recently obtained by the CMS Collaboration.
\section{Acknowledgements}

We thank CERN for the very successful operation of the LHC, as well as the
support staff from our institutions without whom ATLAS could not be
operated efficiently.

We acknowledge the support of ANPCyT, Argentina; YerPhI, Armenia; ARC,
Australia; BMWF, Austria; ANAS, Azerbaijan; SSTC, Belarus; CNPq and FAPESP,
Brazil; NSERC, NRC and CFI, Canada; CERN; CONICYT, Chile; CAS, MOST and
NSFC, China; COLCIENCIAS, Colombia; MSMT CR, MPO CR and VSC CR, Czech
Republic; DNRF, DNSRC and Lundbeck Foundation, Denmark; ARTEMIS, European
Union; IN2P3-CNRS, CEA-DSM/IRFU, France; GNAS, Georgia; BMBF, DFG, HGF, MPG
and AvH Foundation, Germany; GSRT, Greece; ISF, MINERVA, GIF, DIP and
Benoziyo Center, Israel; INFN, Italy; MEXT and JSPS, Japan; CNRST, Morocco;
FOM and NWO, Netherlands; RCN, Norway; MNiSW, Poland; GRICES and FCT,
Portugal; MERYS (MECTS), Romania; MES of Russia and ROSATOM, Russian
Federation; JINR; MSTD, Serbia; MSSR, Slovakia; ARRS and MVZT, Slovenia;
DST/NRF, South Africa; MICINN, Spain; SRC and Wallenberg Foundation,
Sweden; SER, SNSF and Cantons of Bern and Geneva, Switzerland; NSC, Taiwan;
TAEK, Turkey; STFC, the Royal Society and Leverhulme Trust, United Kingdom;
DOE and NSF, United States of America.

The crucial computing support from all WLCG partners is acknowledged
gratefully, in particular from CERN and the ATLAS Tier-1 facilities at
TRIUMF (Canada), NDGF (Denmark, Norway, Sweden), CC-IN2P3 (France),
KIT/GridKA (Germany), INFN-CNAF (Italy), NL-T1 (Netherlands), PIC (Spain),
ASGC (Taiwan), RAL (UK) and BNL (USA) and in the Tier-2 facilities
worldwide.



\bibliographystyle{elsarticle-num}
\bibliography{HiggsToTauTauCOMBPUB}

\onecolumn
\begin{flushleft}
{\Large The ATLAS Collaboration}

\bigskip

G.~Aad$^{\rm 48}$,
B.~Abbott$^{\rm 111}$,
J.~Abdallah$^{\rm 11}$,
A.A.~Abdelalim$^{\rm 49}$,
A.~Abdesselam$^{\rm 118}$,
O.~Abdinov$^{\rm 10}$,
B.~Abi$^{\rm 112}$,
M.~Abolins$^{\rm 88}$,
H.~Abramowicz$^{\rm 153}$,
H.~Abreu$^{\rm 115}$
% ...
\bigskip

% ...
$^{10}$ Institute of Physics, Azerbaijan Academy of Sciences, Baku, Azerbaijan\\
$^{11}$ Institut de F\'isica d'Altes Energies and Universitat Aut\`onoma  de Barcelona and ICREA, Barcelona, Spain\\
% ...
$^{48}$ Fakult\"{a}t f\"{u}r Mathematik und Physik, Albert-Ludwigs-Universit\"{a}t, Freiburg i.Br., Germany\\
$^{49}$ Section de Physique, Universit\'e de Gen\`eve, Geneva, Switzerland\\
% ...
$^{88}$ Department of Physics and Astronomy, Michigan State University, East Lansing MI, United States of America\\
% ...
$^{111}$ Homer L. Dodge Department of Physics and Astronomy, University of Oklahoma, Norman OK, United States of America\\
$^{112}$ Department of Physics, Oklahoma State University, Stillwater OK, United States of America\\
% ...
$^{115}$ LAL, Univ. Paris-Sud and CNRS/IN2P3, Orsay, France\\
% ...
$^{118}$ Department of Physics, Oxford University, Oxford, United Kingdom\\
% ...
$^{153}$ Raymond and Beverly Sackler School of Physics and Astronomy, Tel Aviv University, Tel Aviv, Israel\\
% ...
\end{flushleft}



\end{document}